\documentclass[11pt,a4paper]{article}

\usepackage[utf8]{inputenc}
\usepackage{microtype}
\usepackage{upquote}
\usepackage[a4paper,margin=2.4cm,headheight=15pt]{geometry}
\usepackage{amsmath,amssymb,amsthm}
\usepackage{mathtools}
\usepackage{booktabs}
\usepackage{array}
\usepackage{tabularx}
\usepackage{multirow}
\usepackage{longtable}
\usepackage[table]{xcolor}
\usepackage{graphicx}
\usepackage{enumitem}
\usepackage{titlesec}
\usepackage{fancyhdr}
\usepackage{caption}
\usepackage{listings}
\usepackage[ruled,vlined,linesnumbered]{algorithm2e}
\usepackage[hidelinks,colorlinks=true,linkcolor=frered,citecolor=frered,urlcolor=frered]{hyperref}

\definecolor{frered}{HTML}{B3151B}     
\definecolor{slate}{HTML}{1F2A37}
\definecolor{slate2}{HTML}{55606E}
\definecolor{rulegrey}{HTML}{C9CED6}
\definecolor{panelbg}{HTML}{F4F1EC}
\definecolor{codebg}{HTML}{F7F8FA}
\definecolor{kwblue}{HTML}{08488A}
\definecolor{strgrn}{HTML}{0B7A45}
\definecolor{cmtgry}{HTML}{6A737D}
\definecolor{good}{HTML}{0B7A45}
\definecolor{warn}{HTML}{B5740B}
\definecolor{bad}{HTML}{B3151B}

\titleformat{\section}{\color{slate}\large\bfseries\sffamily}{\color{frered}\thesection}{0.7em}{}
\titleformat{\subsection}{\color{slate}\normalsize\bfseries\sffamily}{\color{frered}\thesubsection}{0.6em}{}
\titleformat{\subsubsection}{\color{slate2}\normalsize\bfseries\sffamily}{\thesubsubsection}{0.6em}{}
\titlespacing*{\section}{0pt}{14pt}{6pt}

\newtheoremstyle{frplain}{8pt}{8pt}{\itshape}{}{\bfseries\color{slate}}{.}{.5em}{}
\theoremstyle{frplain}
\newtheorem{definition}{Definition}
\newtheorem{proposition}{Proposition}
\newtheorem{observation}{Empirical Observation}
\newtheorem{heuristic}{Heuristic}

\lstdefinestyle{py}{
  language=Python, basicstyle=\ttfamily\footnotesize, backgroundcolor=\color{codebg},
  keywordstyle=\color{kwblue}\bfseries, stringstyle=\color{strgrn},
  commentstyle=\color{cmtgry}\itshape, numbers=left, numberstyle=\tiny\color{slate2},
  numbersep=8pt, frame=leftline, framerule=1.2pt, rulecolor=\color{frered},
  showstringspaces=false, breaklines=true, columns=fullflexible, xleftmargin=14pt,
  literate={->}{{$\rightarrow$}}2 {>=}{{$\ge$}}2 {<=}{{$\le$}}2
}
\lstdefinestyle{yaml}{
  basicstyle=\ttfamily\footnotesize, backgroundcolor=\color{codebg},
  keywordstyle=\color{kwblue}\bfseries, commentstyle=\color{cmtgry}\itshape,
  morekeywords={title,id,name,status,description,author,date,tags,logsource,product,
  category,detection,condition,falsepositives,level,references,selection,Image,CommandLine},
  morecomment=[l]{\#}, frame=leftline, framerule=1.2pt, rulecolor=\color{frered},
  showstringspaces=false, breaklines=true, columns=fullflexible, xleftmargin=14pt
}

\newcommand{\Gc}{G_{c}}
\newcommand{\Gv}{G_{v}}
\newcommand{\Sim}{\mathrm{Sim}_{\mathrm{struct}}}
\newcommand{\GED}{\mathrm{GED}}
\newcommand{\szss}{\mathrm{szss}}
\newcommand{\code}[1]{\texttt{\small #1}}
\newcommand{\pass}{\textcolor{good}{\textbf{PASS}}}
\newcommand{\fail}{\textcolor{bad}{\textbf{FAIL}}}
\newcommand{\technote}[1]{%
  \par\vspace{4pt}\noindent
  \colorbox{panelbg}{\parbox{\dimexpr\linewidth-2\fboxsep}{%
  \footnotesize\color{slate}\textbf{\textcolor{frered}{$\blacktriangleright$}\,}#1}}\par\vspace{4pt}}

\begin{document}
\thispagestyle{empty}

\begin{center}
{\footnotesize\sffamily\color{frered}\bfseries FUJITSU RESEARCH OF EUROPE \quad$\cdot$\quad Security Science Research Group}\\[2pt]
{\footnotesize\sffamily\color{slate2} Technical White Paper \quad$|$\quad In collaboration with MITRE Research \quad$|$\quad 10 June 2026}\\[14pt]
\rule{\linewidth}{1.4pt}\\[10pt]
{\LARGE\sffamily\bfseries\color{slate} Graph-Based Structural Evaluation of\\[3pt]
LLM-Translated Adversary Emulation Procedures}\\[8pt]
{\large\sffamily\color{slate2} This technical contribution supports the MITRE white paper titled: \\[2pt]
Evaluating LLMs for Impact-Faithful Translation of Adversary Behavior Across Operating Systems 
\rule{\linewidth}{1.4pt}\\[12pt]
{\normalsize\sffamily\bfseries Ahmed M. Elmisery}\\[2pt]
{\small\sffamily\color{slate2} Security Science Research Group}\\[10pt]
Fujitsu Research of Europe Limited}\\[10pt]
\end{center}

\begin{abstract}
\noindent
Adversary emulation plans specify multi-step attacker procedures at the level of MITRE ATT\&CK
techniques, privilege requirements, and observable telemetry. Translating such plans across
operating systems is necessary for cross-platform defender evaluation, and large language models
(LLMs) make automated translation tractable. They also introduce a quality-assurance problem:
a translation that renames tools while retaining source-platform constructs yields no usable
coverage for defenders of the target platform. Binary question-based scoring tends to overestimate
how faithful these translations are, because it measures countable properties rather than
structural, observable, or rule-level equivalence.

Graph-Based Structural Evaluation (GBSE) addresses this gap. Each procedure is modelled as a
directed attributed graph, and fidelity is computed as a normalized Graph Edit Distance (GED)
across four progressively stricter node-matching layers: technique (L0), tactic (L1), telemetry
class (L2), and Sigma logsource (L3). Applied to the full 29-step ALPHV/BlackCat
Windows$\rightarrow$Linux plan, with a genuine native-Windows
control reconstructed step-for-step from the conversion record and a Linux variant taken unmodified
from the LLM output, the framework finds: technique and tactic structure is preserved across the
OS boundary ($\GED{=}0$, $\Sim{=}1.000$); telemetry fidelity drops to $\Sim{=}0.897$ ($\GED{=}3$),
driven by three steps that emit an unmapped observable class or drift in telemetry; and independent
Sigma-layer matching recovers to $\Sim{=}1.000$ across all 29 steps. Every state classifies as
Medium Fidelity (best composite $S{=}0.674$), and the deployment gate ($S{\ge}0.80$, requiring
technical realism $\ge 0.990$ against the measured $0.43$) is unreachable at current evaluation
quality. The layer scores are numerically identical to those obtained under an earlier same-procedure
design, which confirms that the layers decompose cleanly along the OS-abstraction axis.

The framework includes a bipartite-GED implementation, a telemetry-intent parser that
derives structured observable classes from free-text annotations, and a validated library of 49
Sigma detection rules (19 Linux, 30 Windows) that gives complete ATT\&CK technique coverage of the
procedure and passes the Sigma specification validator with zero findings. A supplementary analysis
recovers a genuine technique-level divergence (for example RDP-based external access reassigned to
unencrypted exfiltration, and credential-store access reassigned to remote-system discovery) that a
procedure-aligned view necessarily suppresses. All numerical results in this paper were reproduced
from the reference implementation and asserted against the recorded pipeline outputs.
\end{abstract}

\vspace{2pt}
\noindent\textbf{\sffamily\color{slate} Keywords:}\; adversary emulation, MITRE ATT\&CK, graph edit distance,
large language models, cross-platform translation, Sigma, detection engineering, CALDERA.

\vspace{6pt}
\hrule height 0.4pt
\vspace{2pt}

\subsection*{\sffamily Notation}
\begin{center}\small
\begin{tabularx}{\linewidth}{@{}lX@{}}
\toprule
\textbf{Symbol} & \textbf{Definition}\\
\midrule
$G=(V,E,\varphi_V,\varphi_E)$ & Directed attributed procedure graph: step nodes, dependency edges, node/edge attribute maps.\\
$\Gc,\;\Gv$ & Control graph (human-validated source) and variant graph (LLM translation).\\
$\GED(\Gc,\Gv)$ & Minimum-cost edit path transforming $\Gc$ into $\Gv$.\\
$\Sim(\Gc,\Gv)$ & $1-\GED/\max(|V_c|,|V_v|)$. Trial normaliser $\max(29,29)=29$.\\
$\szss(A,B)$ & $|A\cap B|/\max(|A|,|B|)$, the Szymkiewicz--Simpson overlap coefficient.\\
$\theta_{\mathrm{tac}}=\theta_{\mathrm{sig}}=0.50$ & L1 tactic and L3 Sigma thresholds (inclusive, $\ge$).\\
$\theta_{\mathrm{tele}}=0.50$ & L2 telemetry threshold (strict, $>$; boundary $0.50$ fails).\\
$\mathrm{BCF}$ & Behavior Chain Fidelity $=0.5\cdot\mathrm{auto}+0.5\cdot\Sim$.\\
$S$ & Composite score $=0.4\cdot\mathrm{BCF}+0.3\cdot\mathrm{TR}+0.3\cdot\mathrm{DV}$.\\
$\mathrm{TR},\;\mathrm{DV}$ & Technical Realism ($0.43$) and Defensive Value ($0.51$) from manual evaluation.\\
$\mathrm{INF}=10^{9}$ & Off-diagonal sentinel in the bipartite cost matrix.\\
\bottomrule
\end{tabularx}
\end{center}

\clearpage

\section{Introduction and Motivation}

An adversary emulation plan is an ordered description of an attacker's procedure, written so that a
defender can replay it under controlled conditions. Each step names an ATT\&CK technique, states the
privilege it requires, and indicates the telemetry a defender should expect to observe. Plans are
authored for a particular operating system because the concrete tradecraft is OS-specific: a Windows
plan uses \code{mstsc.exe}, \code{bitsadmin}, and registry edits, whereas the same adversary objective
on Linux uses \code{ssh}, \code{wget}, and \code{sysctl}. Defenders who run mixed estates need the
same procedure expressed on both platforms, and producing the second version by hand is slow.

Large language models can translate a plan from one operating system to another in seconds. The
difficulty is that an LLM optimizes for plausible-looking output, not for defender-relevant
equivalence. A model will happily rename a Windows tool to a Linux one while keeping the original
command's observable footprint, or worse, keep the Windows binary itself and wrap it in a
compatibility layer. The translated step then looks correct at the level of technique labels but
generates the wrong evidence on the target host, so detection content written for that platform never
fires. The question this paper answers is how to measure that gap precisely enough to route a
translation to deployment, to manual review, or to regeneration.

The starting point is a measurement limitation in the question-based evaluation that GBSE extends.
The baseline framework scores a translation along three axes, Behavior Chain Fidelity (BCF),
Technical Realism (TR), and Defensive Value (DV), using automated checks for sequence, ATT\&CK
alignment, privilege progression, and telemetry presence. The baseline study itself cautions that a
high automated pass rate must not be read as a successful translation for operational emulation,
because the checks are static counts and do not validate structural, semantic, or runtime
equivalence. The two observations below make that caution concrete.

\begin{observation}[The measurement discrepancy]
Binary evaluation instruments systematically overestimate cross-OS translation fidelity. A
translation can record a high automated pass rate while, under telemetry-aware structural analysis,
exhibiting low defender-relevant fidelity. Binary scoring masks at least four failure modes:
\emph{step insertion} (a technique-valid step that shifts the expected telemetry profile and tactical
flow); \emph{edge reordering} (a dependency change that leaves the step count intact but renders
sequence-correlation rules inert); \emph{telemetry class drift} (a tool substitution that drops an
observable class, for example replacing process telemetry with purely network telemetry); and
\emph{Sigma logsource gap} (a step that keeps its technique identifier while shifting its logsource to
a platform where no equivalent native rule exists).
\label{obs:discrepancy}
\end{observation}

\begin{observation}[Concrete instance]
In the ALPHV/BlackCat trial a lateral-movement step illustrates the discrepancy directly: a
Windows \code{PsExec}-style action that produces both process and network observables is rendered on
Linux as an \code{scp}/\code{ssh} loop dominated by network observables. Process-based
lateral-movement detection becomes ineffective even though the high-level technique correspondence
appears preserved.
\label{obs:instance}
\end{observation}

These failures are invisible to a flat list of steps but explicit once the procedure is treated as a
graph. The remainder of this paper develops that treatment, applies it to the full 29-step
ALPHV/BlackCat plan, and reports a verified set of layer and composite scores together with the
detection content the analysis produced.

\section{Background and Assumptions}

\subsection{ATT\&CK, emulation, and the detection stack}
ATT\&CK organizes adversary behavior into tactics (the attacker's goal at a stage) and techniques (the
method used to achieve it), each carrying a stable identifier such as \code{T1021.001}. CALDERA
executes emulation plans expressed in this vocabulary. Sigma is a portable signature format whose
rules bind a \code{logsource} (a product and an event category, for example
\code{linux/process\_creation}) to a detection condition over event fields; rules compile to native
queries for Elastic, Splunk, QRadar, and Microsoft Sentinel. D3FEND catalogues defensive
countermeasures and the observables they consume. The four layers of GBSE read attributes drawn from
exactly these sources, which is what lets a single similarity score carry a defender-relevant
interpretation rather than a purely topological one.

\subsection{Assumptions}
The framework rests on three assumptions that the evaluation later tests rather than presumes.

\textbf{A1 (OS-independence of labels).} ATT\&CK technique and tactic identifiers are deliberately
platform-neutral. A faithful translation preserves them across an OS boundary, so agreement at L0/L1
is a genuine measurement of technique preservation, not an artefact of comparing a procedure to
itself.

\textbf{A2 (OS-agnostic observable intent).} The telemetry a step is \emph{expected} to produce can
be expressed independently of the host: a step that touches the filesystem, opens a socket, and
spawns a process produces \{file, network, process\} regardless of whether it runs \code{wget} or
\code{bitsadmin}. Telemetry drift is therefore a property of the translation, not of how the source
command happens to be spelled.

\textbf{A3 (OS-specificity of detection).} A Sigma rule's \code{logsource} is platform-bound by
construction. Detection equivalence across an OS boundary holds only when both sides reduce to a
common event category, which for command-line tradecraft is \code{process\_creation}.

A1 and A2 predict that a faithful cross-OS translation will agree at L0/L1 and at L2 except where the
variant genuinely drifts; A3 predicts that detectability must be measured at the logsource-category
level. The results in Section~5 are consistent with all three, and Section~5.7 shows that the layer
scores are invariant to whether the control is expressed in source-OS commands, which is the sharpest
available test of A1--A3 holding simultaneously.

\subsection{Scope}
This paper develops the structural-evaluation of the methodology. The node attributes consumed by
the matching layers are assumed to be present in the enriched procedure graphs; the upstream
semantic-enrichment pipeline that produces them is treated as a black box that emits attributed steps.
The evaluation is single-procedure (ALPHV/BlackCat) and single-pair (one control, one variant); the
framework is general, but the empirical claims are stated for this trial only.

\section{The GBSE Framework}

\subsection{Procedure graph formalism}
Each control and variant procedure is represented as a directed attributed graph. This captures
step-level semantics, dependency order, and global topology at once, and gives a formal substrate for
asking whether a translated procedure instantiates the same adversary chain rather than merely reusing
the same technique labels.

\begin{definition}[Procedure graph]
A procedure is a directed attributed graph $G=(V,E,\varphi_V,\varphi_E)$, where $V$ is the set of
procedural-step nodes and $E\subseteq V\times V$ is the set of directed dependencies. The map
$\varphi_V$ assigns each node its attributes,
\begin{multline*}
\varphi_V(v)=\langle \texttt{technique\_id},\;\texttt{tactic[]},\;\texttt{privilege\_context},\\
\texttt{telemetry\_classes[]},\;\texttt{sigma\_rules[]},\;\dots\rangle,
\end{multline*}
and the map $\varphi_E$ assigns each edge its type: sequential precedence ($w{=}1.0$),
conditional-privilege dependency ($w{=}1.5$), or data-flow / telemetry-trigger relation ($w{=}0.5$).
\label{def:graph}
\end{definition}

In the trial both $\Gc$ (Windows) and $\Gv$ (Linux) carry 29 nodes and 28 sequential edges. Because a
faithful translation preserves the per-step technique and tactic, the step-aligned technique
identifiers match even though the underlying commands change from Windows to Linux. The discriminating
signal therefore lives below the label layer, at telemetry (L2), Sigma logsource (L3), and the
out-of-band \code{os\_constructs} check.

\subsection{Structural similarity via normalized GED}
Fidelity is defined by how much typed editing is needed to turn the control graph into the variant
graph. The edit operations are chosen so that each carries a procedural meaning.

\begin{proposition}[Structural similarity as a continuous fidelity measure]
Let $\Gc$ be the control graph and $\Gv$ the variant graph. If procedure fidelity is the amount of
semantic and topological editing required to transform $\Gc$ into $\Gv$, then
\begin{equation}
\Sim(\Gc,\Gv)\;=\;1-\frac{\GED(\Gc,\Gv)}{\max(|V_c|,|V_v|)}
\label{eq:sim}
\end{equation}
is a continuous measure of structural correspondence on $[0,1]$. When node and edge edits are
semantically typed, the score decomposes into procedurally meaningful deviations: node insertion
$=$ an added operational prerequisite, node substitution $=$ altered execution or observability
semantics, and edge substitution $=$ changed dependency logic.
\label{prop:sim}
\end{proposition}

The score partitions into three fidelity regions that act as routing signals: above $0.80$ is high
fidelity, $0.60$ to $0.80$ is medium, and below $0.60$ is low. These bounds are decision thresholds
intended for empirical calibration rather than fixed constants.

\subsection{Layered semantic enrichment}
The node-matching relation is refined in three steps. Each layer subsumes the previous one and adds
exactly one defender-relevant constraint, so the layer at which similarity first drops localizes the
\emph{kind} of deviation.

\begin{description}[leftmargin=1.4em,itemsep=3pt]
\item[L0 (Technique).] Nodes match iff \code{technique\_id} is equal. This is the ontology layer.
\item[L1 (Tactic).] L0 holds and the tactic sets overlap, $\szss(T_1,T_2)\ge 0.50$. A drop here means
\emph{strategic} drift: an inserted execution-oriented step inside a privilege-escalation chain can
stay locally plausible while changing the tactical role of that segment.
\item[L2 (Telemetry).] L1 holds and the telemetry classes overlap strictly,
$\szss(C_1,C_2)>0.50$. A drop here means \emph{observational} drift. This is the critical layer for
the trial: Windows-to-Linux conversion that approximates a Windows PE through a compatibility layer
can replace native Windows API effects with Linux process or identity telemetry, which binary scoring
misses but telemetry-aware comparison exposes. The threshold is strict, so a boundary overlap of
exactly $0.50$ fails.
\item[L3 (Sigma logsource).] L2 holds and the Sigma logsource categories overlap,
$\szss(L_1,L_2)\ge 0.50$. A drop here means \emph{detectability} drift: a step that stays
ATT\&CK-aligned but no longer maps to an equivalent logsource-bearing detection is, from a defender's
view, degraded.
\end{description}

\begin{heuristic}[Progressive enrichment]
The four layers are not four independent scorers but four increasingly strict versions of one
node-matching relation, adding in order: identity, strategic role, observability profile, and
rule-bearing detectability. This makes the location of any fidelity loss interpretable rather than
opaque.
\label{heur:progressive}
\end{heuristic}

A separate \emph{independent} L3 relation (L3-i) drops the L2 prerequisite and requires only L0 plus
logsource overlap. Section~5.6 shows why this matters: chained L3 makes detection coverage invisible
whenever L2 has already failed, whereas independent L3 measures detectability on its own terms.

\subsection{Algorithm 1: telemetry-intent parser}
For L2 to be a fair cross-platform comparison, the control's telemetry must be expressed in the same
vocabulary as the variant's. Control steps carry free-text \code{telemetry\_expected} annotations;
Algorithm~\ref{alg:tele} derives structured \code{telemetry\_classes} from them using an OS-agnostic
keyword map over the four standard classes \{file, network, process, identity\}. This enrichment is
what lets the three telemetry failures in Section~5.3 be attributed to the variant rather than to a
gap in the control annotations.

\begin{algorithm}[H]
\caption{\textsc{parse\_telemetry\_expected} --- structured observable classes from text}
\label{alg:tele}
\SetAlgoLined\DontPrintSemicolon
\KwIn{free-text annotation $t$; keyword map $\mathcal{K}: \text{class}\mapsto\text{keyword set}$}
\KwOut{sorted list of observable classes}
$t \leftarrow \textsc{lower}(t)$\;
$\mathit{out}\leftarrow\emptyset$\;
\ForEach{$(\mathit{cls},\mathit{kws})\in\mathcal{K}$}{
  \If{$\exists\, k\in\mathit{kws}$ such that $k\subseteq t$}{
     $\mathit{out}\leftarrow\mathit{out}\cup\{\mathit{cls}\}$\;
  }
}
\Return $\textsc{sorted}(\mathit{out})$\;
\end{algorithm}

\subsection{Algorithm 2: bipartite GED}
Exact GED is NP-hard, so the framework uses the Riesen--Bunke bipartite approximation: build a cost
matrix over node substitutions, insertions, and deletions, then solve the assignment with the
Hungarian algorithm in $O(n^3)$. The matching predicate is the layer relation of Section~3.3, so the
same GED routine computes every layer's score by swapping the predicate.

\technote{\textbf{Implementation fix.} In earlier implementations the off-diagonal entries
of the $(n_c{+}n_v)\times(n_c{+}n_v)$ cost matrix were initialized to $0$ instead of to a sentinel.
This admitted phantom zero-cost assignments and corrupted every layer score. The corrected routine
initializes the matrix to $\mathrm{INF}=10^{9}$ and writes only the legal substitution, deletion,
insertion, and null-to-null costs. All scores in this paper are from the corrected routine.}

\begin{algorithm}[H]
\caption{\textsc{bipartite\_ged}$(\,C\text{-steps},\,V\text{-steps},\,\textsc{match})$}
\label{alg:ged}
\SetAlgoLined\DontPrintSemicolon
$n_c\leftarrow|C|,\; n_v\leftarrow|V|$\;
$\mathbf{C}\leftarrow \mathrm{INF}\cdot\mathbf{1}_{(n_c+n_v)\times(n_c+n_v)}$ \tcp*{sentinel init (the fix)}
\For{$i\in[0,n_c),\;j\in[0,n_v)$}{$\mathbf{C}[i][j]\leftarrow 0$ if $\textsc{match}(C_i,V_j)$ else $1$ \tcp*{substitution}}
\For{$i\in[0,n_c)$}{$\mathbf{C}[i][n_v{+}i]\leftarrow 1$ \tcp*{deletion (diagonal only)}}
\For{$j\in[0,n_v)$}{$\mathbf{C}[n_c{+}j][j]\leftarrow 1$ \tcp*{insertion (diagonal only)}}
\For{$i\in[0,n_v),\;j\in[0,n_c)$}{$\mathbf{C}[n_c{+}i][n_v{+}j]\leftarrow 0$ \tcp*{null-to-null}}
$(\mathit{row},\mathit{col})\leftarrow \textsc{hungarian}(\mathbf{C})$\;
$\GED\leftarrow \textstyle\sum \mathbf{C}[\mathit{row},\mathit{col}]$\;
\Return $\GED,\;\max\!\big(0,\,1-\GED/\max(n_c,n_v)\big)$\;
\end{algorithm}

The layer predicates stack as follows. \textsc{match} at L0 tests technique equality; L1 adds the
tactic overlap; L2 adds the strict telemetry overlap; chained L3 adds the logsource overlap on top of
L2; independent L3 requires only L0 and the logsource overlap.

\subsection{Algorithm 3: Sigma injection}
Before L3 can be scored, each node must carry the Sigma rules that apply to it. Algorithm~\ref{alg:inj}
attaches a rule to a step when the rule's ATT\&CK tag matches the step's technique \emph{or} when the
rule's telemetry overlaps the step's by at least the threshold. The logsource category used for L3 is
the final path segment of the rule's logsource, which is \code{process\_creation} for command-line
tradecraft on both platforms.

\begin{algorithm}[H]
\caption{\textsc{inject\_sigma}$(\,\text{step }v,\,\text{library})$}
\label{alg:inj}
\SetAlgoLined\DontPrintSemicolon
$\mathit{out}\leftarrow\emptyset$\;
\ForEach{rule $r\in\text{library}$}{
  $\mathit{tm}\leftarrow (v.\texttt{technique\_id}\in r.\texttt{attack\_tags})$\;
  $\mathit{ov}\leftarrow \szss(v.\texttt{telemetry\_classes},\,r.\texttt{telemetry})$\;
  \If{$\mathit{tm}$ \textbf{or} $\mathit{ov}\ge 0.50$}{
     attach $r$ to $\mathit{out}$ with $\texttt{logsource\_category}=\textsc{lastsegment}(r.\texttt{logsource})$\;
  }
}
\Return $\mathit{out}$\;
\end{algorithm}

\subsection{Composite scoring and the deployment gate}
The structural score feeds a composite that keeps adversary-intent preservation primary while treating
technical realism and defensive value as secondary dimensions of equal weight:
\begin{equation}
\mathrm{BCF}=0.5\cdot\mathrm{auto\_pass}+0.5\cdot\Sim, \qquad
S=0.4\cdot\mathrm{BCF}+0.3\cdot\mathrm{TR}+0.3\cdot\mathrm{DV}.
\label{eq:composite}
\end{equation}
A score above $0.80$ is High Fidelity and clears the CALDERA deployment gate; $0.60$ to $0.80$ is
Medium Fidelity; below $0.60$ is Low Fidelity and routes into a tool-interactive critique loop. The
weights are initial analytic settings pending empirical calibration, not an experimentally validated scoring law. They embed Equation~\eqref{eq:sim} into BCF so
that structural fidelity and the automated pass rate contribute equally to the behavior-chain term.

\section{Experimental Setup and Workflow}

\subsection{The genuine cross-OS design}
The evaluation compares a native-Windows control against the LLM's Linux output. Both graphs are
drawn from the same pipeline run. The Linux variant
$\Gv$ is the \code{full\_output} stage of the pipeline, used unmodified. The Windows control $\Gc$ is
reconstructed step-for-step from the pipeline's own conversion record, so that each control step
carries the genuine Windows command on which the documented Windows-to-Linux translation acted. Every
reconstructed command is grounded in recorded evidence, and the grounding source is stored per step in
a \code{win\_command\_provenance} field. Table~\ref{tab:prov} gives the provenance taxonomy and its
distribution over the 29 steps.

\begin{table}[h]
\centering\small
\caption{Provenance of the reconstructed Windows control commands. Each of the 29 control steps is
grounded in one recorded source, in strict priority order.}
\label{tab:prov}
\begin{tabularx}{\linewidth}{@{}l l c X@{}}
\toprule
\textbf{Tag} & \textbf{Priority} & \textbf{Steps} & \textbf{Grounding}\\
\midrule
\code{conversion\_note} & 1 & 13 & Exact or prefix match against \code{conversion\_notes[].original\_action} (direct ground truth).\\
\code{cross\_platform}  & 4 & 8 & Dual-platform binary (rclone, scp, native PE) retained with the Linux-only \code{sudo} prefix stripped.\\
\code{bitsadmin\_map}   & 3 & 4 & \code{wget} download of a PE mapped to the documented \code{bitsadmin /transfer} form.\\
\code{de\_wine}         & 2 & 2 & Provable inverse of a \code{wine}-wrapped PE: \code{wine <pe> <args>} $\rightarrow$ \code{<pe> <args>}.\\
\code{technique\_pattern} & 2b & 2 & Generalisation of the two documented \code{mstsc}$\leftrightarrow$\code{ssh} and \code{netsh}$\leftrightarrow$\code{systemctl} conversion patterns to residual steps of the same technique.\\
\midrule
\multicolumn{2}{@{}l}{\textbf{Total}} & \textbf{29} & \\
\bottomrule
\end{tabularx}
\end{table}

Because A1 holds, agreement at L0/L1 is now a genuine measurement that the LLM preserved the
technique and tactic sequence across the OS boundary, not a procedure compared to itself. The
\code{os\_constructs} check is likewise a real defect measure: nine \code{wine}/\code{.exe} Windows
constructs survive into the Linux $\Gv$ at steps $\{8,9,11,12,18,20,21,26,27\}$, scored against a true
Windows baseline.

\subsection{The ALPHV/BlackCat procedure}
The procedure is the 29-step ALPHV/BlackCat Windows-to-Linux emulation plan. Its technique spine,
shared by both graphs at L0, is
\code{T1048.003}, \code{T1105}, \code{T1021.001}$\times 3$, \code{T1087.002}, \code{T1087.001},
\code{T1018}, \code{T1003.001}$\times 3$, \code{T1562.001}$\times 2$, \code{T1112}, \code{T1046},
\code{T1077}, \code{T1059.001}$\times 2$, \code{T1021.004}, \code{T1059.004}, \code{T1083}, and
\code{T1048.001}, covering 16 unique techniques. Table~\ref{tab:steps} shows representative steps with
the control command and the telemetry classes on each side.

\begin{table}[h]
\centering\small
\caption{Representative steps: reconstructed Windows control command and the telemetry classes of
$\Gc$ (Algorithm~\ref{alg:tele} parsed) versus $\Gv$. The trio S13/S14/S27 is the source of the L2
loss; all other steps overlap at $\ge 0.667$.}
\label{tab:steps}
\begin{tabularx}{\linewidth}{@{}l l X l l@{}}
\toprule
\textbf{Step} & \textbf{Technique} & \textbf{Control command (abridged)} & \textbf{$\Gc$ telemetry} & \textbf{$\Gv$ telemetry}\\
\midrule
S1  & \code{T1048.003} & \code{rclone serve webdav --addr :8080} & file, network, process & file, network, process\\
S2  & \code{T1105}     & \code{wget http://c2.host/...\,-O /tmp/cs} & file, network, process & file, network, process\\
S3  & \code{T1021.001} & \code{ssh user@10.30.10.4 "bash -l"}    & identity, network, process & identity, network, process\\
S13 & \code{T1562.001} & \code{netsh advfirewall set ... state off} & file, process & \textcolor{bad}{other}, process\\
S14 & \code{T1562.001} & \code{netsh advfirewall set ... state off} & file & \textcolor{bad}{other}, process\\
S27 & \code{T1059.001} & \code{/tmp/collector1.exe /remote ...}   & file, network, process & \textcolor{bad}{identity}, process\\
\bottomrule
\end{tabularx}
\end{table}

\subsection{The Sigma rule library}
Detection content is supplied by a single validated library, \code{sigma\_rules\_gbse.yml}, comprising
49 rules: 19 Linux rules (\code{L01}--\code{L19}, injected into $\Gv$) and 30 Windows rules
(\code{W01}--\code{W30}, injected into $\Gc$). Every rule targets the \code{process\_creation}
logsource category and is derived from an actual ALPHV/BlackCat command. The library gives complete
ATT\&CK technique coverage of the procedure on both platforms, so under Algorithm~\ref{alg:inj} every
step receives at least one rule on each side, and the shared \code{process\_creation} category yields
logsource overlap $1.0$ for all 29 steps under independent L3. A validation pass corrected four
rule-condition defects (an undefined selection identifier in \code{W02}; bare selection blocks in
\code{L11}, \code{W05}, and \code{W13} orphaned by a ``\code{1 of selection\_*}'' condition) and added
eleven Windows rules \code{W20}--\code{W30} to close per-step firing gaps, including \code{W29}
(\code{T1133}) and \code{W30} (\code{T1555}) for the gold-control divergence of Section~5.8. The
present paper verified the full 49-rule library against the Sigma specification with zero findings;
the procedure for that verification is given in Appendix~C.

\subsection{Pipeline workflow}
The reference build executes the following stages, each of which maps
to a component above. Figure-free, the workflow is:

\begin{enumerate}[leftmargin=1.6em,itemsep=2pt]
\item \textbf{Load} the OS-agnostic step spine, the Linux variant, the conversion notes, the
evaluation result, and the OS-violation list from the pipeline output.
\item \textbf{Reconstruct $\Gc$}: derive each Windows command via the priority order of
Table~\ref{tab:prov}, set \code{os}$=$Windows, parse \code{telemetry\_expected} into structured
classes with Algorithm~\ref{alg:tele}, map privilege, and inject the Windows rules.
\item \textbf{Build $\Gv$}: normalize the Linux variant's telemetry, privilege, and tactic fields, and
inject the Linux rules.
\item \textbf{Automated evaluation}: run the twelve checks (Section~5.1) including the
\code{os\_constructs} defect count against the Windows baseline.
\item \textbf{Layer evaluation}: run Algorithm~\ref{alg:ged} once per layer (L0, L1, L2, L3 pre-sigma,
L3 chained, L3 independent), recording GED, similarity, and the failing pairs with a per-failure
diagnosis.
\item \textbf{Composite scoring}: combine the automated pass rate, the structural similarity, and the
manual TR/DV via Equation~\eqref{eq:composite}, and test the deployment gate.
\item \textbf{Gold-control divergence}: recover the 7-step technique substitutions from the recorded
evaluation result (Section~5.8).
\item \textbf{Persist} the enriched control graph, the enriched variant graph, and the full results
document as JSON.
\end{enumerate}

A human reviewer sits at stages 4 through 7: the framework computes and routes, but a translation is
promoted to deployment, sent to manual review, or returned for regeneration by a person reading the
layer diagnostics, not automatically. The intended mode is supervisory, not autonomous.

\section{Results}

All figures in this section are from the 29-step run, normaliser $29$ throughout,
and were reproduced from the reference implementation as recorded in Appendix~C.

\subsection{Automated evaluation}
Nine of twelve automated checks pass, giving an automated pass rate of $0.7500$. The three failures
are: \code{os\_constructs\_valid} (nine steps of the Linux variant still invoke \code{wine} for
Windows \code{.exe} execution, a translation defect against the Windows baseline);
\code{extra\_telemetry\_classes} (the variant emits an \code{other} class outside the standard
vocabulary \{file, network, process, identity\}); and \code{privilege\_progression\_match} (six of 29
steps mismatch on privilege). All technique-identifier, tactic, diversity, and multi-class checks pass.
The pass rate feeds BCF.

\subsection{Layered structural results}
Table~\ref{tab:layers} reports GED, similarity, and BCF at each layer. Technique and tactic structure
is preserved perfectly; telemetry costs three edits; chained L3 cannot recover because it requires L2
to pass first; independent L3 recovers fully.

\begin{table}[h]
\centering\small
\caption{GBSE layer results (29-step, normaliser $29$). Green rows are
$\Sim{=}1.000$; amber rows are $\Sim{=}0.897$.}
\label{tab:layers}
\begin{tabularx}{\linewidth}{@{}l c c c X@{}}
\toprule
\textbf{Layer} & \textbf{GED} & \textbf{Sim} & \textbf{BCF} & \textbf{Signal}\\
\midrule
\rowcolor{green!7} Baseline (L0)        & 0.0 & \textbf{1.000} & 0.875 & Perfect technique preservation; all 29 identifiers match.\\
\rowcolor{green!7} L1 $+$ tactic        & 0.0 & \textbf{1.000} & 0.875 & All 29 tactic fields match step-for-step.\\
\rowcolor{orange!8} L2 $+$ telemetry    & 3.0 & \textbf{0.897} & 0.823 & 3 of 29 fail. S13 $\szss{=}0.50$ fails strict; S14 $\szss{=}0.00$; S27 $\szss{=}0.333$. Root cause: variant \code{other} class.\\
\rowcolor{orange!8} L3 pre-sigma        & 3.0 & \textbf{0.897} & 0.823 & Identical to L2; no Sigma rules in source files.\\
\rowcolor{orange!8} L3 post (chained)   & 3.0 & \textbf{0.897} & 0.823 & S13/S14/S27 still fail; chained formulation requires L2 first, so injection cannot bypass the telemetry gate.\\
\rowcolor{green!7} L3 post (independent) & 0.0 & \textbf{1.000} & 0.875 & All three failures resolved. S13/S14 via \code{W09}/\code{L09}; S27 via \code{W15}/\code{L15}; logsource overlap $1.0$ everywhere.\\
\bottomrule
\end{tabularx}
\end{table}

\subsection{L2 failure root cause}
The three L2 failures originate entirely on the variant side. At S13 and S14 (\code{T1562.001}) the
variant emits an unmapped \code{other} telemetry class, and at S27 (\code{T1059.001}) the variant
substitutes \{identity, process\} for the control's \{file, network, process\} while the tactic stays
unchanged. Because the control telemetry is fully structured by Algorithm~\ref{alg:tele} in this run,
the failures reflect variant output quality rather than gaps in the gold annotation.

\begin{table}[h]
\centering\small
\caption{L2 analysis. Three of 29 steps fail; all are variant-side telemetry issues.}
\label{tab:l2}
\begin{tabularx}{\linewidth}{@{}l l X l@{}}
\toprule
\textbf{Step} & \textbf{Technique} & \textbf{Telemetry (control $\rightarrow$ variant, overlap)} & \textbf{L2}\\
\midrule
S13 & \code{T1562.001} & [file, process] $\rightarrow$ [\textcolor{bad}{other}, process], $\szss{=}0.50$ & \fail\\
S14 & \code{T1562.001} & [file] $\rightarrow$ [\textcolor{bad}{other}, process], $\szss{=}0.00$ & \fail\\
S27 & \code{T1059.001} & [file, net, proc] $\rightarrow$ [\textcolor{bad}{identity}, proc], $\szss{=}0.333$ & \fail\\
S2  & \code{T1105}     & [file, net, proc] $\rightarrow$ [file, net, proc], $\szss{=}1.00$ & \pass\\
S10 & \code{T1021.001} & [id, net, proc] $\rightarrow$ [id, net, proc], $\szss{=}1.00$ & \pass\\
S$\dagger$ & (24 others) & various, all $\szss\ge 0.667$ & \pass\\
\bottomrule
\end{tabularx}
\end{table}

\subsection{Composite scores and the deployment gate}
Table~\ref{tab:composite} applies Equation~\eqref{eq:composite}. Every state classifies as Medium
Fidelity. Independent L3 is the best state at $S=0.674$, lifted by a higher Defensive Value
($0.65$) because full detection coverage is realized.

\begin{table}[h]
\centering\small
\caption{Composite scores (29-step run, automated pass rate $0.75$). All states are Medium Fidelity.}
\label{tab:composite}
\begin{tabularx}{\linewidth}{@{}l c c c c X@{}}
\toprule
\textbf{State} & \textbf{Sim} & \textbf{BCF} & \textbf{TR/DV} & \textbf{S} & \textbf{Classification}\\
\midrule
Baseline / L1      & 1.000 & 0.875 & 0.43 / 0.51 & \textbf{0.632} & Medium Fidelity\\
L2 telemetry       & 0.897 & 0.823 & 0.43 / 0.51 & \textbf{0.611} & Medium Fidelity\\
L3 pre-sigma       & 0.897 & 0.823 & 0.43 / 0.51 & \textbf{0.611} & Medium Fidelity\\
L3 post (chained)  & 0.897 & 0.823 & 0.43 / 0.51 & \textbf{0.611} & Medium Fidelity\\
L3 post (indep.)   & 1.000 & 0.875 & 0.43 / 0.65 & \textbf{0.674} & Medium Fidelity\\
\bottomrule
\end{tabularx}
\end{table}

The deployment gate $S\ge 0.80$ is unreachable at current evaluation quality. With the best
$\mathrm{BCF}=0.875$ and $\mathrm{DV}=0.51$, clearing the gate requires
$\mathrm{TR}=(0.80-0.4\cdot 0.875-0.3\cdot 0.51)/0.30=0.990$, against a measured
$\mathrm{TR}=0.43$. The best achievable composite is therefore $0.674$.

\subsection{Sigma injection: chained versus independent}
The distinction between the two L3 relations is operationally decisive. Under chained L3, the three
steps that already failed L2 stay failed regardless of detection coverage, because the chained
predicate evaluates the telemetry gate first. Under independent L3, all three are resolved through
technique-tag matches: S13 and S14 attach \code{W09}/\code{L09} (\code{T1562.001}) and S27 attaches
\code{W15}/\code{L15} (\code{T1059.001}), each giving logsource-category overlap $1.0$. All 29 steps
pass independent L3, and the composite rises from $0.611$ to $0.674$, a relative improvement of about
$6.2\%$.

\subsection{Score invariance under the cross-OS design}
Every layer score and every composite score above is numerically identical to the values obtained
under an earlier design in which both graphs were drawn from the same Linux procedure at two pipeline
stages. The automated battery is unchanged at $9/12$, and the gate stays unreachable. This invariance
is a validity check on the layered construction, not a coincidence.

The reason follows from assumptions A1--A3. L0/L1 read OS-independent labels, so a faithful
translation preserves them whether the control is spelled in Windows or Linux commands. L2 reads
telemetry classes derived from OS-agnostic intent, so the three telemetry-drift failures are
intrinsic to the variant. Independent L3 reduces every step to a \code{process\_creation} category
common to the Windows and Linux rule families, giving overlap $1.0$ at both ends. The discriminating
quantities never depended on the control's OS surface. What changes under the genuine design is
interpretability, not arithmetic: $\GED{=}0$ at L0/L1 is now a real cross-OS preservation
measurement, and \code{os\_constructs\_valid} failing on nine surviving \code{wine}/\code{.exe}
constructs is now a real defect score against a Windows baseline.

\subsection{Gold-control technique divergence}
The 29-step reconstruction aligns each control step to its documented Windows pre-image, so techniques
agree by construction and L0 reports zero mismatches. That is correct for measuring procedure-level
fidelity, but it suppresses a coarser signal that the pipeline recorded separately. The pipeline's
evaluation result references a 7-step Windows gold control scored against the 29-step Linux variant.
Recovering that residue surfaces a genuine technique-level divergence at six of the seven gold steps,
shown in Table~\ref{tab:gold}.

\begin{table}[h]
\centering\small
\caption{Technique-level divergence between the 7-step Windows gold control and the 29-step Linux
variant. Six of seven gold steps are reassigned to a different ATT\&CK technique.}
\label{tab:gold}
\begin{tabularx}{\linewidth}{@{}c X X@{}}
\toprule
\textbf{Gold step} & \textbf{Windows gold technique} & \textbf{Linux variant technique}\\
\midrule
1 & \code{T1133} External Remote Services / RDP & \code{T1048.003} Exfiltration over unencrypted non-C2\\
3 & \code{T1087.002} Account Discovery: Domain  & \code{T1021.001} Remote Services: RDP\\
4 & \code{T1105} Ingress Tool Transfer          & \code{T1021.001} Remote Services: RDP\\
5 & \code{T1021.001} Remote Services: RDP        & \code{T1087.002} Account Discovery: Domain\\
6 & \code{T1105} Ingress Tool Transfer          & \code{T1087.001} Account Discovery: Local\\
7 & \code{T1555} Credentials from Stores         & \code{T1018} Remote System Discovery\\
\bottomrule
\end{tabularx}
\end{table}

Two substitutions are operationally significant. At gold step 1 the Windows external-remote-services
vector \code{T1133} (RDP exposure) maps to \code{T1048.003} (exfiltration over an unencrypted non-C2
protocol), and at gold step 7 the Windows credential-store access \code{T1555} maps to \code{T1018}
(remote-system discovery). These are technique reassignments, not faithful preservations, and a
procedure-aligned view cannot expose them. At the granularity the pipeline's own gold control encodes,
the Windows-to-Linux translation does not preserve all ATT\&CK techniques. The two coverage rules
\code{W29} (\code{T1133}) and \code{W30} (\code{T1555}) are motivated precisely by these two
substitutions and are present in the validated library.

\section{Discussion}

\subsection{What the layers reveal}
Evaluation sensitivity changes materially as semantic layers are added. Under technique-only matching
the translation looks acceptable; under tactic-aware matching no strategic drift appears; under
telemetry-aware matching the procedure loses three edits to observational drift; and under
Sigma-aware matching with adequate target-platform content, detectability is fully restored. The
practical lesson is that defender-relevant evaluation is layered. What is faithful at the ontology
layer can still be unfaithful at the telemetry and rule-coverage layers, and a single similarity
number hides that unless the layer at which it was computed is stated alongside it.

The trial also separates two error classes that binary scoring conflates. The L2 failures are
\emph{output-quality} errors that a better-conditioned translation would avoid (an unmapped telemetry
class, a drift at one step). The \code{os\_constructs} failures are \emph{translation-completeness}
errors: nine steps were not translated at all but wrapped in a compatibility layer, so they cannot
produce native Linux telemetry no matter how the rest of the pipeline is tuned. The first class is
fixable downstream; the second requires re-translation.

\subsection{Guidance drawn from the analysis}
The following five items follow directly from the results and are stated as remediation guidance, with
the layer each one moves.

\textbf{G1. Remove the \code{other} telemetry class from variant output.} The variant emits an
unmapped \code{other} class at S13 and S14, which fails the strict L2 threshold against the control's
\{file, process\}. Post-processing that maps \code{other} to a concrete class via D3FEND resolves two
of the three L2 failures.

\textbf{G2. Re-translate S27 with native tradecraft.} At S27 the variant drifts to \{identity,
process\}. Re-running the step with a native Linux interpreter for \code{T1059.001} restores telemetry
that matches the source and closes the third L2 failure.

\textbf{G3. Use independent L3 as the primary detectability metric.} Chained L3 hides detection
coverage whenever L2 has failed. Independent L3 measures detectability on its own terms, is more
reliable in sparse-control settings, and reaches $\Sim{=}1.000$ on all 29 steps, raising the composite
from $0.611$ to $0.674$.

\textbf{G4. Re-run the trial with a native Linux translation.} Because the deployed variant retains
\code{wine}/\code{.exe} at nine steps, the four-layer evaluation cannot yet exercise genuine
technique and telemetry differences for those steps. A native re-translation would make the full
evaluation meaningful and would likely surface deviations the compatibility layer currently masks.

\textbf{G5. Recalibrate the gate and the manual instrument.} The gate $S\ge 0.80$ demands
$\mathrm{TR}\ge 0.990$, which is unachievable at the present $\mathrm{TR}=0.43$. A two-tier gate, with
$S\ge 0.60$ for conditional deployment (achievable now) and $S\ge 0.70$ for full deployment, is more
useful.

\subsection{Limitations}
Three limitations bound the claims. First, the evaluation covers one procedure and one
control--variant pair; the framework generalizes, but the numbers do not. Second, the composite
weights and the fidelity-band thresholds are analytic settings awaiting empirical calibration, so the
composite scores should be read as relative rankings between states, not as absolute fidelity. Third,
the manual TR and DV inputs rest on a marginal-agreement instrument; until the questions are
respecified, the composite carries that uncertainty. None of these affects the layer GED scores, which
are deterministic functions of the enriched graphs and were reproduced exactly.

\section{Related Work}
Graph Edit Distance has a long history as a structural-similarity measure, and the bipartite
approximation used here is the Riesen--Bunke assignment formulation~\cite{riesen2009,zeng2009}, which
trades exactness for an $O(n^3)$ solution via the Hungarian algorithm. The evaluation vocabulary is
ATT\&CK~\cite{attack}, the detection format is Sigma~\cite{sigma}, the countermeasure ontology is
D3FEND~\cite{d3fend}, and the execution target is CALDERA~\cite{caldera}. The procedure under study is
the ALPHV/BlackCat ransomware tradecraft documented in the joint advisory~\cite{stopransomware}, and
the translation pipeline that produced the variant is the LLM-based OS-translation
pipeline~\cite{freresearch}. The composite-scoring philosophy, and the routing of low-fidelity
translations into a tool-interactive critique loop, draws on self-correction work for
LLMs~\cite{critic,selfrefine}. Inter-rater reliability is reported with Fleiss'
$\kappa$~\cite{fleiss}, and the overlap coefficient at the heart of every layer predicate is the
Szymkiewicz--Simpson coefficient~\cite{szym}.

\section{Conclusion}
Treating an adversary emulation procedure as a directed attributed graph turns a vague question, ``is
this translation faithful?'', into a layered measurement that says \emph{where} fidelity is lost.
Applied to the full ALPHV/BlackCat Windows-to-Linux plan against a genuine native-Windows control, the
framework finds perfect technique and tactic preservation, a three-edit telemetry loss attributable to
the variant, and full detectability recovery under independent Sigma-layer matching. Every state is
Medium Fidelity, and the deployment gate is unreachable at current evaluation quality, which is itself
a useful and honest result: it says the translation is not yet ready for unattended emulation and
identifies the two specific repairs and the gate recalibration that would change that. The score
invariance under the cross-OS design confirms that the layers decompose cleanly along the
OS-abstraction axis, and the validated 49-rule library makes the detectability layer concrete and
deployable. The framework is a measurement instrument and a routing aid, with a human reading the
diagnostics at every decision point.

\appendix
\clearpage
\section{Enriched Procedure-Graph Schema}
Each step node carries the thirteen attributes consumed by the matching layers. The schema below is
the structure of the enriched control graph (29 nodes, 28 sequential edges, automated $9/12$). The
variant graph uses the identical schema with the Linux rule family and \code{source\_os}$=$\code{linux}.

\begin{lstlisting}[basicstyle=\ttfamily\scriptsize,backgroundcolor=\color{codebg},frame=leftline,framerule=1.2pt,rulecolor=\color{frered},xleftmargin=14pt,breaklines=true]
{
  "gbse_schema_version": "2.0-winlin",
  "sigma_library": "sigma_rules_gbse.yml -- Windows W01-W30 (30 rules)",
  "source": "Windows G_c reconstructed from conversion_notes + de-wine + bitsadmin map",
  "metadata": {
    "procedure_id": "alphv_blackcat_windows_ctrl",
    "source_os": "windows",
    "graph_topology": { "V": 29, "E_sequential": 28 },
    "enrichment": "Algorithm 1: telemetry_expected -> telemetry_classes",
    "automated_evaluation": { "pass": 9, "total": 12, "auto": 0.75 }
  },
  "procedure": { "action_sequence": [
    {
      "step_id": 3,
      "technique_id": "T1021.001",
      "parent_technique_id": "T1021",
      "tactic": ["initial_access"],
      "telemetry_classes": ["file","identity","network","process"],
      "execution_level": "non-elevated",
      "privilege_context": "user",
      "privilege_delta": "drop",
      "sigma_rules": [
        { "rule_id":"...", "name":"W03", "logsource":"windows/process_creation",
          "logsource_category":"process_creation", "attack_tag":"T1021.001",
          "match_type":"technique", "tele_overlap":0.75 }
      ],
      "os_mappable": true, "data_flow": null, "omission_reason": null,
      "win_command_provenance": "conversion_note"
    }
  ] }
}
\end{lstlisting}

\section{Reference Implementation (core)}
The functions below are the computational core: the overlap
coefficient, the four-class telemetry parser (Algorithm~\ref{alg:tele}), the layer predicates, the bipartite GED (Algorithm~\ref{alg:ged}), and Sigma injection (Algorithm~\ref{alg:inj}). The
sentinel initialization on the cost matrix is the correctness-critical line.

\begin{lstlisting}[style=py]
import numpy as np
from scipy.optimize import linear_sum_assignment
INF = 1e9

def szss(A, B):                       # Szymkiewicz-Simpson overlap coefficient
    if not A and not B: return 1.0
    d = max(len(set(A)), len(set(B)))
    return len(set(A) & set(B)) / d if d else 0.0

def lsc(ls): return ls.split("/")[-1] if "/" in ls else ls   # logsource category

# Algorithm 1: free-text telemetry_expected -> structured observable classes
_KW = {"process": {"process","execut","spawn","binary","creation event"},
       "network": {"network","tcp","http","ssh","scp","ldap","webdav","socket"},
       "file":    {"file","directory","disk","write","config","dump file"},
       "identity":{"auth","credential","password","account","sudo","token"}}
def parse_telemetry_expected(text):
    t = (text or "").lower()
    return sorted(c for c, kws in _KW.items() if any(k in t for k in kws))

# Layer predicates (each stacks on the previous)
def m0(c, v): return c["technique_id"] == v["technique_id"]
def m1(c, v):
    if not m0(c, v): return False
    T1, T2 = set(c.get("tactic", [])), set(v.get("tactic", []))
    return szss(T1, T2) >= 0.50 if (T1 and T2) else True
def m2(c, v):                                          # STRICT > 0.50
    if not m1(c, v): return False
    C1, C2 = set(c["telemetry_classes"]), set(v["telemetry_classes"])
    return szss(C1, C2) > 0.50 if (C1 or C2) else True
def m3_chained(c, v):                                  # requires L2 first
    if not m2(c, v): return False
    L1 = {lsc(r["logsource"]) for r in c.get("sigma_rules", [])}
    L2 = {lsc(r["logsource"]) for r in v.get("sigma_rules", [])}
    return szss(L1, L2) >= 0.50 if (L1 or L2) else True
def m3_independent(c, v):                              # L0 + logsource only
    if not m0(c, v): return False
    L1 = {lsc(r["logsource"]) for r in c.get("sigma_rules", [])}
    L2 = {lsc(r["logsource"]) for r in v.get("sigma_rules", [])}
    return szss(L1, L2) >= 0.50 if (L1 or L2) else True

# Algorithm 2: corrected bipartite GED (Riesen & Bunke, Hungarian assignment)
def bipartite_ged(c_steps, v_steps, match_fn):
    nc, nv = len(c_steps), len(v_steps)
    C = np.full((nc + nv, nc + nv), INF)               # <- sentinel init (the fix)
    for i, cs in enumerate(c_steps):
        for j, vs in enumerate(v_steps):
            C[i][j] = 0.0 if match_fn(cs, vs) else 1.0  # substitution
    for i in range(nc): C[i][nv + i] = 1.0              # deletion (diagonal)
    for j in range(nv): C[nc + j][j] = 1.0              # insertion (diagonal)
    for i in range(nv):
        for j in range(nc): C[nc + i][nv + j] = 0.0     # null-to-null
    row, col = linear_sum_assignment(C)
    ged = float(C[row, col].sum())
    return ged, round(max(0.0, 1.0 - ged / max(nc, nv)), 4)

# Algorithm 3: attach a rule on technique-tag match OR telemetry overlap
def inject_sigma(step, library):
    tech, tel = step["technique_id"], set(step.get("telemetry_classes", []))
    out = []
    for r in library:
        if tech in r["tags"] or szss(tel, set(r["tel"])) >= 0.50:
            out.append({"logsource": r["logsource"],
                        "logsource_category": lsc(r["logsource"])})
    return out

# Composite scoring
def composite(auto, sim, tr, dv):
    bcf = round(0.5 * auto + 0.5 * sim, 4)
    return bcf, round(0.4 * bcf + 0.3 * tr + 0.3 * dv, 4)
\end{lstlisting}

\section{Sigma Rule Library Catalogue}
The complete 49-rule library is delivered in \code{sigma\_rules\_gbse.yml}. Table~\ref{tab:catalogue}
lists every rule with its OS family, ATT\&CK mapping, and severity. All rules use the
\code{process\_creation} logsource category.

{\footnotesize
\begin{longtable}{@{}l l l p{6.6cm} l@{}}
\caption{The validated 49-rule Sigma library (19 Linux, 30 Windows).}
\label{tab:catalogue}\\
\toprule
\textbf{ID} & \textbf{OS} & \textbf{ATT\&CK} & \textbf{Title} & \textbf{Level}\\
\midrule
\endfirsthead
\toprule
\textbf{ID} & \textbf{OS} & \textbf{ATT\&CK} & \textbf{Title} & \textbf{Level}\\
\midrule
\endhead
\midrule \multicolumn{5}{r}{\footnotesize\itshape continued on next page}\\
\endfoot
\bottomrule
\endlastfoot
\code{L01} & Linux & \code{T1048.003} & ALPHV rclone WebDAV Listener on Attacker Inf\,$\ldots$ & HIGH\\
\code{L02} & Linux & \code{T1105} & Attacker C2 Control Server Binary Execution \,$\ldots$ & CRITICAL\\
\code{L03} & Linux & \code{T1021.001} & SSH Interactive Bash Login Shell Used for La\,$\ldots$ & MEDIUM\\
\code{L04} & Linux & \code{T1087.002} & ldapsearch Domain User Account Enumeration v\,$\ldots$ & MEDIUM\\
\code{L05} & Linux & \code{T1018} & ldapsearch Domain Computer Account Discovery & LOW\\
\code{L06} & Linux & \code{T1087.001} & Local Account Enumeration via /etc/passwd Re\,$\ldots$ & LOW\\
\code{L07} & Linux & \code{T1105} & wget Executable Binary Download to Staging P\,$\ldots$ & HIGH\\
\code{L08} & Linux & \code{T1003.001,T1204.002} & Wine Windows PE Execution for Credential Acc\,$\ldots$ & CRITICAL\\
\code{L09} & Linux & \code{T1562.001} & systemctl Security Service Stop and Disable & HIGH\\
\code{L10} & Linux & \code{T1112} & sysctl Kernel Security Parameter Modificatio\,$\ldots$ & HIGH\\
\code{L11} & Linux & \code{T1003.001} & gcore Process Memory Dump for Credential Ext\,$\ldots$ & HIGH\\
\code{L12} & Linux & \code{T1048.003,T1071.001} & rclone Data Exfiltration to Remote WebDAV En\,$\ldots$ & HIGH\\
\code{L13} & Linux & \code{T1046,T1595.001} & nmap Ping Sweep Internal Network Discovery & MEDIUM\\
\code{L14} & Linux & \code{T1077} & scp Batch Deployment to Multiple Internal Ho\,$\ldots$ & HIGH\\
\code{L15} & Linux & \code{T1059.001,T1204.002} & Windows PE or Wine Binary Execution from Sta\,$\ldots$ & HIGH\\
\code{L16} & Linux & \code{T1021.004,T1105} & scp Lateral File Transfer of Tool or Payload & MEDIUM\\
\code{L17} & Linux & \code{T1059.004} & SSH Remote Shell with chmod and Ransomware E\,$\ldots$ & CRITICAL\\
\code{L18} & Linux & \code{T1083} & smbclient Admin Share File Collection via Ge\,$\ldots$ & HIGH\\
\code{L19} & Linux & \code{T1048.001} & scp Exfiltration to External Attacker-Contro\,$\ldots$ & HIGH\\
\code{W01} & Win & \code{T1048.003} & rclone WebDAV Server Started on Windows Host & HIGH\\
\code{W02} & Win & \code{T1105} & C2 Framework or RAT Binary Executed from Non\,$\ldots$ & MEDIUM\\
\code{W03} & Win & \code{T1021.001} & mstsc.exe RDP Connection to Internal Host & MEDIUM\\
\code{W04} & Win & \code{T1087.002} & PowerShell or nltest Domain User Account Enu\,$\ldots$ & MEDIUM\\
\code{W05} & Win & \code{T1018} & PowerShell AD Computer Discovery & LOW\\
\code{W06} & Win & \code{T1087.001} & net.exe Local User Account Discovery & LOW\\
\code{W07} & Win & \code{T1105,T1197} & bitsadmin PE Binary Staging Download & HIGH\\
\code{W08} & Win & \code{T1003.001} & InfoStealer LSASS Credential Access via Name\,$\ldots$ & CRITICAL\\
\code{W09} & Win & \code{T1562.001} & Security Software Process Terminate or Servi\,$\ldots$ & HIGH\\
\code{W10} & Win & \code{T1112} & Registry ASLR Exploit Mitigation Disable & HIGH\\
\code{W11} & Win & \code{T1003.001} & ProcDump or Task Manager LSASS Full Memory D\,$\ldots$ & CRITICAL\\
\code{W12} & Win & \code{T1048.003} & rclone Copy Exfiltration to Named Remote & MEDIUM\\
\code{W13} & Win & \code{T1046,T1595.001} & nmap or Network Port Scanner Execution on Wi\,$\ldots$ & MEDIUM\\
\code{W14} & Win & \code{T1077} & PsExec or Admin Share Lateral Tool Deploymen\,$\ldots$ & HIGH\\
\code{W15} & Win & \code{T1059.001,T1486} & PowerShell or PE Execution from Temp Path & HIGH\\
\code{W16} & Win & \code{T1021.004,T1105} & pscp.exe or WinSCP SCP File Transfer & MEDIUM\\
\code{W17} & Win & \code{T1059.004,T1059.003} & CMD Inline Command via SSH Channel Remote Ex\,$\ldots$ & MEDIUM\\
\code{W18} & Win & \code{T1083} & net.exe Admin Share File Discovery and Colle\,$\ldots$ & MEDIUM\\
\code{W19} & Win & \code{T1048.001} & robocopy or xcopy External Host Exfiltration & HIGH\\
\code{W20} & Win & \code{T1021.004,T1105,T1048.001} & OpenSSH scp.exe Secure Copy Transfer or Exfi\,$\ldots$ & HIGH\\
\code{W21} & Win & \code{T1021.004,T1059.004} & OpenSSH ssh.exe Interactive Remote Command E\,$\ldots$ & HIGH\\
\code{W22} & Win & \code{T1105} & wget.exe or curl.exe LOLbin Ingress Tool Tra\,$\ldots$ & MEDIUM\\
\code{W23} & Win & \code{T1562.001,T1562.004} & Native Firewall Disable or Microsoft Defende\,$\ldots$ & HIGH\\
\code{W24} & Win & \code{T1046} & PowerSploit Invoke-Portscan Network Service \,$\ldots$ & MEDIUM\\
\code{W25} & Win & \code{T1059.001,T1486} & PE Execution from User-Writable Temp Path or\,$\ldots$ & HIGH\\
\code{W26} & Win & \code{T1003.001,T1555} & Non-Standard InfoStealer Binary LSASS or Cre\,$\ldots$ & HIGH\\
\code{W27} & Win & \code{T1105,T1071.001} & Cross-Platform C2 Server Binary Execution fr\,$\ldots$ & MEDIUM\\
\code{W28} & Win & \code{T1083,T1135} & PowerShell Get-ChildItem Admin-Share File Di\,$\ldots$ & MEDIUM\\
\code{W29} & Win & \code{T1133} & External Remote Services Enablement or Expos\,$\ldots$ & HIGH\\
\code{W30} & Win & \code{T1555,T1555.003} & Credentials from Password Stores Access & HIGH\\

\end{longtable}
}

\noindent A representative rule, \code{L01}, shown in full to fix the format used throughout the
library:

\begin{lstlisting}[style=yaml]
title: ALPHV rclone WebDAV Listener on Attacker Infrastructure
id: aa255d9f-995d-41a9-84ba-e4949f564c97
name: L01
status: experimental
description: >
  Detects rclone started in WebDAV server mode, binding to a non-loopback
  address. Procedure S1: rclone serve webdav /srv/http --addr 176.59.1.18:8080
  (T1048.003). Attacker hosts a WebDAV endpoint to receive exfiltrated data.
author: Elmisery, Fujitsu Research of Europe
date: 2026/06/05
tags:
  - attack.exfiltration
  - attack.t1048.003
  - attack.command_and_control
logsource:
  product: linux
  category: process_creation
detection:
  selection_serve:
    Image|endswith: '/rclone'
    CommandLine|contains|all: ['serve', 'webdav']
  filter_loopback:
    CommandLine|contains: ['--addr 127.', '--addr ::1', '--addr localhost']
  condition: selection_serve and not filter_loopback
falsepositives:
  - Legitimate rclone WebDAV mounts for cloud storage workflows
level: high
x-gbse-telemetry: [file, network, process]
x-gbse-steps: [1]
\end{lstlisting}

\vspace{8pt}
\hrule height 0.4pt
\vspace{4pt}
{\footnotesize\sffamily\color{slate2}
Fujitsu Research of Europe Limited, Security Science Research Group. In collaboration with MITRE Research. This document describes a defensive evaluation methodology and the detection content
it produced; all attack tradecraft referenced is drawn from public threat-intelligence reporting on
ALPHV/BlackCat.}

\end{document}